\documentclass[aps,amsmath,amssymb,pre,twocolumn,showpacs,superscriptaddress]{revtex4}
\usepackage{graphicx}
\usepackage{lineno}
\usepackage{natbib}

\begin{document}

\title{Establishing a direct connection between detrended fluctuation analysis and Fourier analysis}
\author{Ken Kiyono}
\affiliation{Graduate School of Engineering Science, Osaka University, 1-3 Machikaneyama-cho, Toyonaka, Osaka 560-8531, Japan}

\date{\today}

\begin{abstract}
To understand methodological features of the detrended fluctuation analysis (DFA) using a higher-order polynomial fitting, we establish the direct connection between DFA and Fourier analysis. Based on an exact calculation of the single-frequency response of the DFA, the following facts are shown analytically: (1) in the analysis of stochastic processes exhibiting a power-law scaling of the power spectral density (PSD), $S(f) \sim f^{-\beta}$, a higher-order detrending in the DFA has no adverse effect in the estimation of the DFA scaling exponent $\alpha$, which satisfies the scaling relation $\alpha = (\beta+1)/2$; (2) the upper limit of the scaling exponents detectable by the DFA depends on the order of polynomial fit used in the DFA, and is bounded by $m + 1$, where $m$ is the order of the polynomial fit; (3) the relation between the time scale in the DFA and the corresponding frequency in the PSD are distorted depending on both the order of the DFA and the frequency dependence of the PSD. We can improve the scale distortion by introducing the corrected time scale in the DFA corresponding to the inverse of the frequency scale in the PSD. In addition, our analytical approach makes it possible to characterize variants of the DFA using different types of detrending. As an application, properties of the detrending moving average algorithm (DMA) are discussed. 
\end{abstract}


\maketitle

\section{Introduction}
Long-range correlations and fractal scaling behavior have been observed in a remarkably wide variety of systems, such as physical \cite{kantelhardt1999phases,vandewalle1999non}, geophysical \cite{telesca2004time,lenton2012early,witt2013quantification}, biological \cite{656,peng1993long,hausdorff1996fractal,K05,bartsch2005statistical,ivanov2009levels}, 
and economic systems \cite{mantegna1999introduction,alvarez2008short,wang2011detrended}. In the study of the time series data observed in such systems, the power spectral analysis is a well-established methodological framework \cite{percival1993spectral,hamilton1994time}. By estimating the slope of the log-log scaled power spectral density (PSD), a wide range of scaling behavior can be characterized. However, it has been pointed out that the power spectral analysis may provide spurious detection of scaling behavior caused by nonstationarity in time series, such as embedded trends and heterogeneous statistical properties \cite{peng1994mosaic,peng1995quantification,hu2001effect,chen2002effect,chen2005effect}. To obtain a more accurate estimate of the scaling exponent, the detrended fluctuation analysis (DFA) has been proposed \cite{peng1994mosaic,peng1995quantification} and has become a widely used method \cite{bartsch2005statistical,ivanov2009levels,matic2015objective,choi2015reliability,varela2015long,rhea2015interpretation}. In this method, local trends in the time series are eliminated by least-squares polynomial fitting. If $m$th-order polynomials are employed in the DFA, it is referred to as $m$th-order DFA or DFA$m$. In addition, variants of the DFA using different types of detrending methods have been proposed \cite{alessio2002second,carbone2004analysis,alvarez2005detrending,chianca2005fourier,xu2005quantifying,bashan2008comparison,qian2011modified,arianos2011self}. 
The statistical performance and the superiority of the DFAs have been shown by a number of numerical studies \cite{kantelhardt2001detecting,bashan2008comparison,bryce2012revisiting}. The effects of nonstationarity, nonlinear filters and extreme data loss on the DFA scaling behavior have also been studied systematically \cite{hu2001effect,chen2002effect,chen2005effect,ma2010effect}. 

However, the mathematical basis of the DFA has not been well established. To date, analytical studies on the DFA have been mostly limited to the first order case \cite{taqqu1995estimators,heneghan2000establishing,talkner2000power,willson2002relationship,willson2003direct,arianos2007detrending,bardet2008asymptotic}. For instance, Taqqu {\it et al}. derived the direct link between the Hurst exponent $H$ of fractional Brownian motion and the scaling exponent estimated by DFA1 \cite{taqqu1995estimators}. Furthermore, several researchers have analytically studied the relation between DFA1 and power spectral analysis to a limited extent \cite{heneghan2000establishing,talkner2000power,willson2002relationship,willson2003direct}. However, there are few analytical arguments on higher-order DFAs. 

In this work, we attempt to achieve a deeper understanding of the methodological features of higher-order DFAs by introducing an analytical approach using the single-frequency response of the DFA. Based on an exact calculation of the single-frequency response function and the assumption of stochastic time series with weak trends, 
the direct connection between higher-order DFAs and Fourier analysis can be derived. In previous studies using numerical experiments, a limitation of the detectable scaling exponent in the DFA has been empirically found \cite{hu2001effect,xu2005quantifying}. In addition, a deviation of the crossover position in the DFA from the corresponding frequency in the PSD was also empirically found \cite{kantelhardt2001detecting}. Our approach can provide clear mathematical reasons for these properties and a guiding principle to improve the DFA methodology. 

The organization of this paper is as follows. In Sec.~\ref{sec:DFA}, we review the DFA method. In Sec. \ref{sec:decomp}, we mention the basic principle of our approach based on the Fourier decomposition. In Sec.~\ref{sec:single_freq}, by considering a single-frequency component, we derive the single-frequency response of DFA. In Sec.~\ref{sec:relation}, using this derived response, we show the direct connection between higher-order DFAs and Fourier analysis. In Sec.~\ref{sec:scaling}, we derive the scaling relation between the scaling exponent estimated by higher-order DFAs and that by the PSD. In Sec.~\ref{sec:ub}, we derive the limitation of the scaling exponent detectable by the DFA. In Sec.~\ref{sec:dist}, the distortion between the time scale in DFA and frequency scale in PSD are studied. Finally, Sec.~\ref{sec:summary} provides a summary of our results, and discusses possible applications of our method.

\begin{figure}[tb]
       \begin{center}
               \includegraphics[width = 0.55\linewidth]{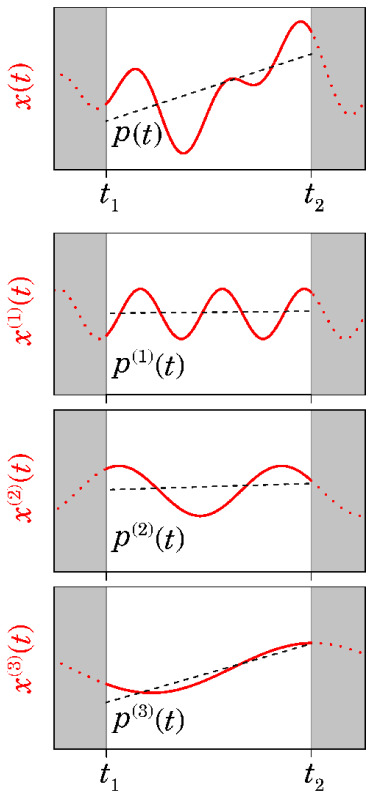}
               \caption{(Color online) Decomposition of least squares fitting. The time series $x(t)$ (top) is given by the sum of three frequency components (three bottom), $x^{(1)}(t)+x^{(2)}(t)+x^{(3)}(t)$. Black dashed lines show the least squares line of each time series. The least squares line $p(t)$ coincides with $p^{(1)}(t)+p^{(2)}(t)+p^{(3)}(t)$. }
               \label{fig:fig1}
       \end{center}
\end{figure}

\section{Detrended fluctuation analysis (DFA)} \label{sec:DFA}

Here, we briefly review the DFA \cite{peng1994mosaic,peng1995quantification}. The standard procedure of the DFA is the following: 1) Starting with 
a time series $\{x_i\}_{i=0}^{N-1}$ of length $N$, the $\{x_i\}$ is integrated after subtracting the mean from each data point: 
\begin{equation}
y_k = \sum_{i=0}^{k-1} \left(x_i - \left\langle x \right\rangle \right), \label{eq:integ_x}
\end{equation}
where $\left\langle x \right\rangle$ denotes the sample mean of $\{x_i\}$. 
 2) The integrated time series $\{y_k\}_{k=1}^{N}$ is divided into equal-sized, non-overlapping segments of length $s$. 3) In each segment, the mean-square-deviation from the least squares polynomial fit of order $m$ is calculated. The mean-square-deviations are then averaged over all segments and its square root $F(s)$, referred to as the fluctuation function, is calculated: 
\begin{equation}
F(s) = \left[\frac{1}{\lfloor N/s \rfloor} \sum_{l=1}^{\lfloor N/s \rfloor} \left\{\frac{1}{s} \sum_{k=(l-1)s+1}^{l s} \left( y_k - p^{(l)}_{k} \right)^2 \right\} \right]^{1/2}, \label{eq:Fs}
\end{equation}
where $\lfloor \cdot \rfloor$ is the floor function, and $p^{(l)}_{k}$ is the least-squares fitting polynomial in the $l$th segment. Steps (2) and (3) are repeated over multiple time scales (window sizes) to characterize the relationship between $F(s)$ and $s$. A linear relationship on a log-log plot of $F(s)$ as function of $s$ indicates the power-law scaling range, in which the fluctuations can be characterized by a scaling exponent $\alpha$, the slope of the linear relation between $\log F(s)$ and $\log s$. 

It is known that the scaling exponent $\alpha$ over large time scales is related to the power spectrum exponent $\beta$ by $\alpha = (\beta + 1)/2$ ($\alpha > 0$), where the PSD of $\{x_i\}$ is assumed as $S(f) \sim f^{-\beta}$. Therefore, a white noise time series with $\beta = 0$ is characterized by $\alpha = 0.5$; a long-range correlated time series with $0 < \beta < 1$ is indicated by $0.5 < \alpha < 1$; and a long-range anti-correlated time series with $-1 < \beta < 0$ is indicated by $0 < \alpha < 0.5$. However, the analytical derivation of the scaling relation between higher-order DFAs and the PSD has not been reported. 

\begin{figure}[t]
       \begin{center}
               \includegraphics[width = 0.75\linewidth]{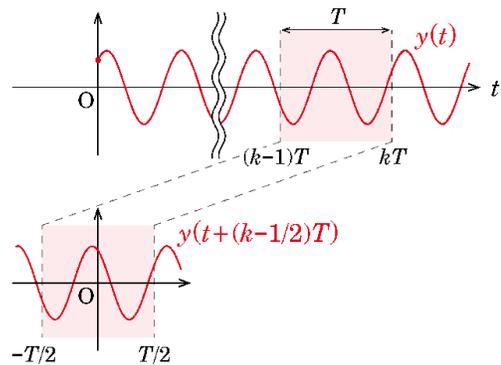}
               \caption{(Color online) Parallel shift of a single frequency component $y(t)$.}
               \label{fig:shift}
       \end{center}
\end{figure}

\section{Decomposition of least squares polynomial and mean square deviation}\label{sec:decomp}

Before deriving the relation between higher-order DFA and Fourier analysis, we mention the basic principle of our approach. A similar approach has been used by Willson {\it et al} to study the relationship between DFA1 and PSD \cite{willson2002relationship,willson2003direct}. 

Here, let us consider continuous-time signals, and assume that a time series $x(t)$ in the range $\left[t_1, t_2\right]$ is expressed by the summation of a set of basis functions $\{ x^{(1)} (t), \cdots, x^{(m)} (t) \}$ [Fig.~\ref{fig:fig1}]:
\begin{equation}
x(t) = \sum_{i=1}^{m} x^{(i)} (t). 
\end{equation}
In this case, if we denote a least squares polynomial fit of $x(t)$ and $\{x^{(i)} (t) \}$ in the range $\left[t_1, t_2\right]$ by $p(t)$ and $\{ p^{(i)} (t)\}$, respectively, the following relation holds (see Fig.~\ref{fig:fig1} and Appendix A for details):
\begin{equation}
p(t) = \sum_{i=1}^{m} p^{(i)} (t). 
\end{equation}
Furthermore, if we consider the square deviation from the least-squares polynomial, we obtain
\begin{widetext}
\begin{eqnarray}
\int_{t_1}^{t_2} \!\!\!\!\!\!& \!&\!\!\!\!\!\! \left\{x(t) - p(x) \right\}^2 \, dt = \int_{t_1}^{t_2} \! \left\{\sum_{i=1}^{m} \left( x^{(i)} (t) - p^{(i)} (x) \right) \right\}^2 \, dt \nonumber \\ 
 &=& \sum_{i=1}^{m} \int_{t_1}^{t_2} \!\! \left\{x^{(i)} (t) - p^{(i)} (x) \right\}^2 \, dt + 2 \sum_{i=1}^{m-1}  \sum_{j= i + 1}^{m}  \int_{t_1}^{t_2} \!\! \left\{x^{(i)} (t) - p^{(i)} (x) \right\} \left\{x^{(j)} (t) - p^{(j)} (x) \right\} \, dt. \nonumber \\ \label{F2decomp}
\end{eqnarray}
\end{widetext}
If the cross terms in Eq.~(\ref{F2decomp}) are negligible, the square deviation of $x(t)$ can be evaluated by the summation of the square deviations of the basis functions $\{x^{(i)} (t) \}$. Note that, in the least-squares fitting, the following always holds: 
\begin{equation}
\int_{t_1}^{t_2} \! \left\{x^{(i)} (t) - p^{(i)} (x) \right\} \, dt = 0 \quad (i = 1, 2, \cdots, m).
\end{equation}
Thus, if values of $x^{(i)} (t) - p^{(i)} (x)$ and $x^{(j)} (t) - p^{(j)} (x)$ ($i \neq j$) are distributed independently of each other, the expected value of the cross term is zero. 

Based on the above facts, we will consider the frequency components of a time series obtained by Fourier transform. In the case of the Fourier decomposition, if the phase differences between different frequency components are uniformly distributed on $[0, 2 \pi]$, the influence of cross-products in Eq.~(\ref{F2decomp}) are expected to be negligible. In stochastic processes displaying power-law spectra, such as fractional Brownian motion and fractional Gaussian noise \cite{mandelbrot1968fractional,flandrin1989spectrum,molz1997fractional,li2006rigorous,li2009fractal}, this condition is expected to hold. However, if a time series includes strong deterministic trends, the influence of cross terms in Eq.~(\ref{F2decomp}) is not negligible. In such a case, our approach is not applicable.

\section{Single-frequency response function of DFA} \label{sec:single_freq}

To derive the single-frequency response of DFA, we consider a single frequency component as a continuous-time signal, 
\begin{equation}
x(t) = A \cos \left(2 \pi f t + \theta \right), \label{eq:x}
\end{equation}
The integrated signal is given by
\begin{equation}
\int_{0}^{t} \! x(\tau) \, d \tau = \frac{A \sin \left(2 \pi f t + \theta \right)}{2 \pi f} - \frac{A \sin \theta }{2 \pi f} \label{eq:integ}
\end{equation}
Because the constant term in Eq.~(\ref{eq:integ}) does not affect the fluctuation function $F(s)$ [Eq.~(\ref{eq:Fs})] in DFA, we neglect the second term and study the integrated signal as 
\begin{equation}
y(t) = \frac{A}{2 \pi f} \sin \left(2 \pi f t + \theta \right). \label{yt}
\end{equation}

\begin{figure}[h]
       \begin{center}
               \includegraphics[width = 1\linewidth]{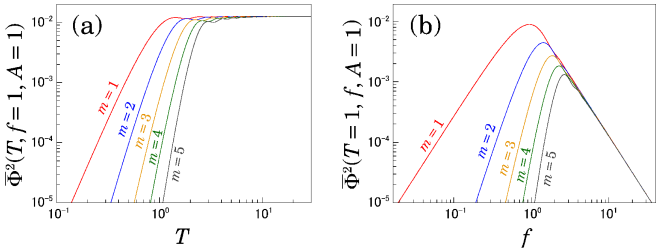}
               \caption{(Color online) Single-frequency response functions $\overline{\Phi}^2 (T, f, A)$ of $m$th-order DFAs ($m = 1, 2, \cdots, 5$). (a) $\overline{\Phi}^2 (T, f=1, A=1)$ versus $T$. (c) $\overline{\Phi}^2 (T=1, f, A=1)$ versus $f$. }
               \label{fig:phi2}
       \end{center}
\end{figure}

\begin{figure}[bh]
       \begin{center}
               \includegraphics[width = 1\linewidth]{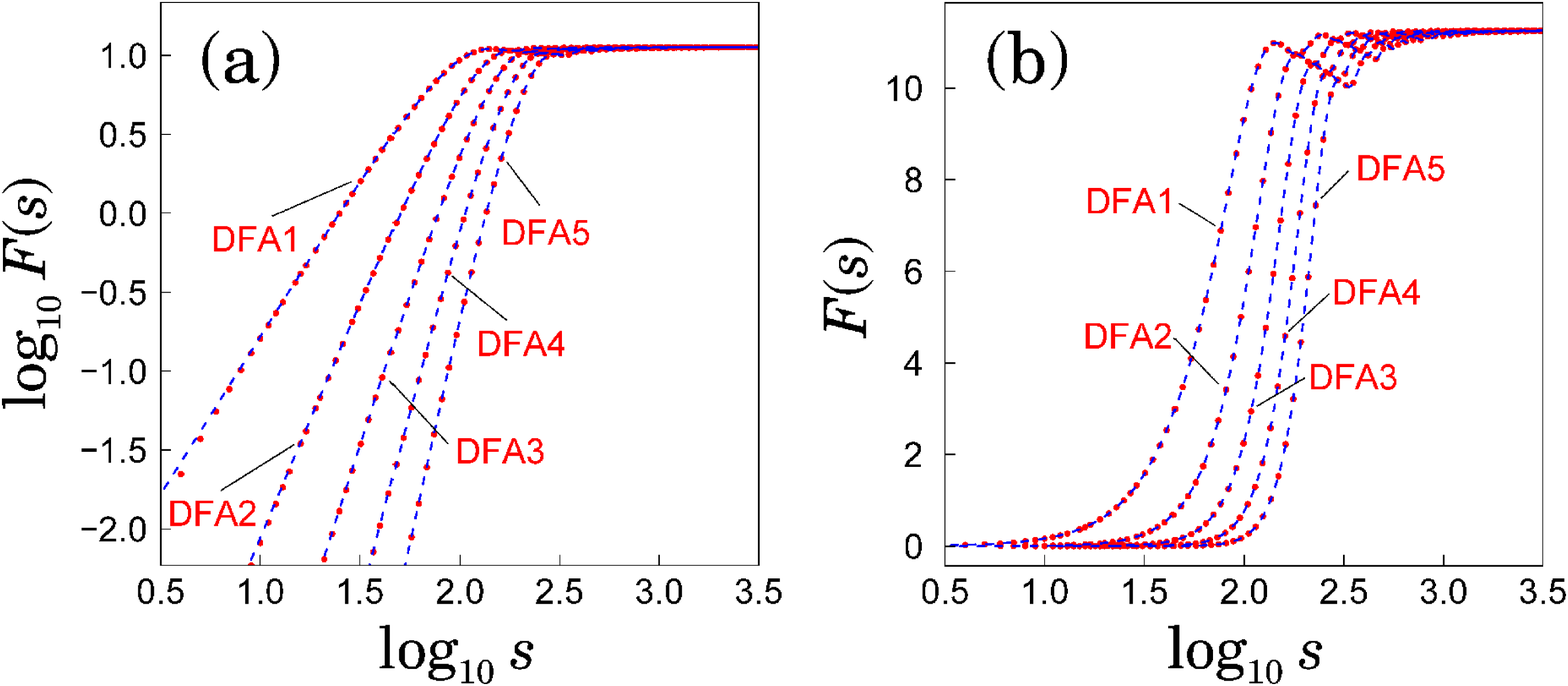}
               \caption{(Color online) Comparison between the DFA results (dots) and the analytical predictions by $\overline{\Phi} (T=s, f=1/100, A=1)$ (dashed lines). The $F(s)$ are obtained by analyzing a time series of a single-frequency component, $x_i = \cos \left(2 \pi i/100 \right)$ ($i=0, 1, \cdots, 10^7-1$), using $m$th-order DFAs ($m = 1, 2, \cdots, 5$). }
               \label{fig:dfa_cos}
       \end{center}
\end{figure}

To obtain the fluctuation function $F(s)$, we have to calculate the mean square deviation from the least squares polynomial in each interval $\left[(k-1) T, k T \right]$, where $T$ is the window length, and $k$ denotes the number of the window. To simplify the calculation, we shift the interval from $\left[(k-1) T, k T \right]$ to $\left[- T/2, T/2 \right]$, as shown in Fig.~\ref{fig:shift}. The mean square deviation of the signal $y(t)$ [Eq.~(\ref{yt})] from its least squares polynomial can be straightforwardly calculated by the following procedure. The least squares polynomial of degree $m$ is obtained by minimizing the following function,
\begin{widetext}
\begin{equation}
I(\{a_0, a_1, \cdots, a_m \}) = \int_{- T/2}^{T/2} \! \left\{\frac{A}{2 \pi f} \sin \left(2 \pi f t + \tilde{\theta} \right) - \sum_{i=0}^m a_i \, t^i \right\}^2 \, dt, 
\end{equation}
\end{widetext}
where $\tilde{\theta} = \theta + \pi f T (2 k -1)$. 
The coefficients $\{a_i \}$ of the polynomial are determined by solving the following equations:
\begin{equation}
\frac{\partial I(\{a_i \})}{\partial a_j} = 0, 
\end{equation}
where $i, j = 0,1,\cdots, m$. After the determination of the coefficients $\{a_i \}$, the mean square deviation $\Phi^2$ is given by 
\begin{widetext}
\begin{equation}
\Phi^2 (T, f, A, \tilde{\theta}) = \frac{1}{T}  \int_{- T/2}^{T/2} \! \left\{\frac{A}{2 \pi f} \sin \left(2 \pi f t + \tilde{\theta} \right) - \sum_{i=0}^m a_i \, t^i \right\}^2 \, dt. \label{eq:Phi2}
\end{equation}
\end{widetext}
Moreover, if the phase $\tilde{\theta}$ in Eq.~(\ref{eq:Phi2}) is averaged out, $\Phi^2$ can be approximated by
\begin{equation}
\overline{\Phi}^2 (T, f, A) = \frac{1}{2 \pi} \int_{0}^{2 \pi} \Phi^2 (T, f, A, \theta)\, d \theta, \label{eq:Phi2ave}
\end{equation}
where we refer to $\overline{\Phi}^2$ as the single-frequency response function. 
The analytical formulas of $\Phi^2$ and $\overline{\Phi}^2$ are summarized in Appendix B. 
The plots of $\overline{\Phi}^2$ up to fifth-order detrending are shown in Fig.~\ref{fig:phi2}. When we analyze a single-frequency component, the square root of $\overline{\Phi}^2 (T=s, f, A)$ provides the analytical prediction of $F(s)$ in the DFA, as shown in Fig.~\ref{fig:dfa_cos}. However, in the range $\log_{10} s < 1.0$ in Fig.~\ref{fig:dfa_cos} (a), very small deviations of the estimated $F(s)$ by the DFA from the analytical prediction by $\overline{\Phi}$ are seen. As will be shown in the next subsection, these deviations come from the continuous-time approximation of the discrete time series.

\subsection{Effect of discrete time sampling}

Under the continuous-time approximation, we can analytically derive the single-frequency response function $\overline{\Phi}^2$ of higher-order DFA. However, in the actual time series analysis, we have to deal with discrete-time signals. Therefore, it is necessary to take the effect of the discrete time sampling into account.

\begin{figure}[t]
       \begin{center}
               \includegraphics[width = 0.55\linewidth]{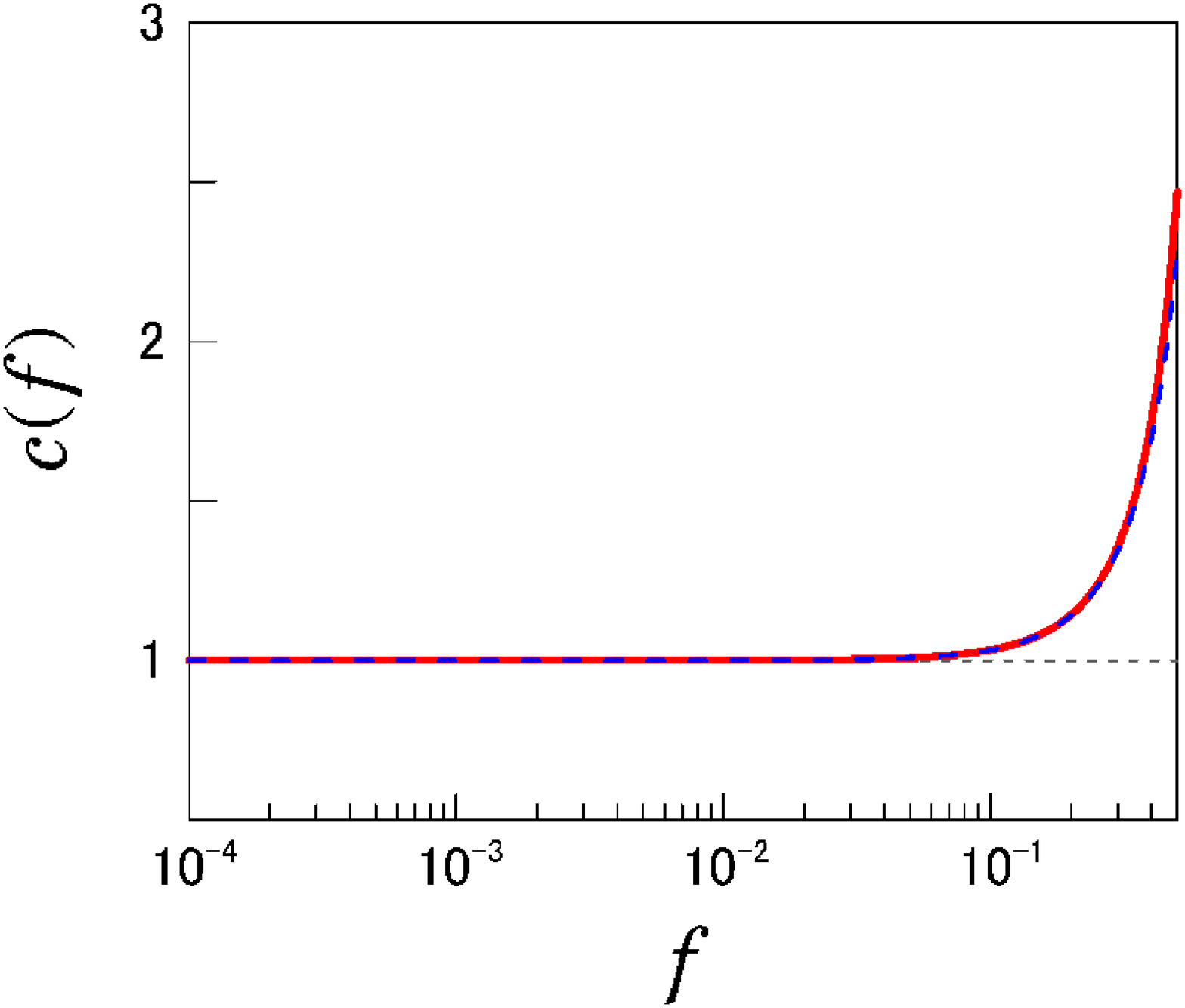}
               \caption{(Color online) Discretization effect evaluated by Eq.~(\ref{eq:c}) (solid line). Blue dashed lines show the approximation by $\exp ( \pi^2 f^2/3)$. }
               \label{fig:c}
       \end{center}
\end{figure}

To study this, let us consider discrete time series $\{x_i\}$,
\begin{equation}
x_i = A \cos \left( 2 \pi f i + \theta \right),  \label{fig:x_i}
\end{equation}
where $i= 0, 1, \cdots, N-1$. The integrated series is given by
\begin{equation}
y_k = \sum_{i=0}^{k-1} x_i = \frac{A \cos \left\{ \pi f (k-1) + \theta \right\} \sin (\pi f k) }{\sin (\pi f)},
\end{equation}
where $k = 1, 2, \cdots, N$. When the window size $s$ is sufficiently long such that the trend in each window is approximated by a constant function, the mean square deviation of $\{y_k\}$ around the baseline is given by
\begin{eqnarray}
\overline{\Phi}_{\rm d}^2 (f, A) &=& \frac{1}{n} \sum_{k=1}^{n} y_k^2 - \left\{ \frac{1}{n} \sum_{k=1}^{n} y_k \right\}^2 \\
&=& \frac{A^2}{8 \sin^2 (\pi f)}.
\end{eqnarray}
On the other hand, the corresponding quantity in the continuous case [Eq.~(\ref{eq:x})] is given 
by taking the limit $T \to \infty$ in Eq.~(\ref{eq:Phi2ave}), 
\begin{eqnarray}
\overline{\Phi}_{\rm c}^2 (f, A) &=& \lim_{T \to \infty} \overline{\Phi}^2 (T, f, A) \\
&=& \frac{A^2}{8 (\pi f)^2}.
\end{eqnarray}
Therefore, the discretization effect can be evaluated by their ratio
\begin{eqnarray}
c(f) &=& \frac{\overline{\Phi}_{\rm d}^2 (f, A)}{\overline{\Phi}_{\rm c}^2 (f, A)}=\frac{(\pi f)^2}{\sin^2 (\pi f)} \label{eq:c} \\ 
&\approx& \exp \frac{\pi^2 f^2}{3} \quad {\rm for}\ f < \frac{1}{2}.  
\end{eqnarray}
As shown in Fig.~\ref{fig:c}, the discretization effect is remarkably strong for the high frequency range $f > 10^{-1}$ and rapidly decreases as the frequency decreases. Therefore, the discretization effect could be negligible, if the high frequency components ($f > 10^{-1}$) are not dominant in the time series. 

\begin{figure}[t]
       \begin{center}
               \includegraphics[width = 0.6\linewidth]{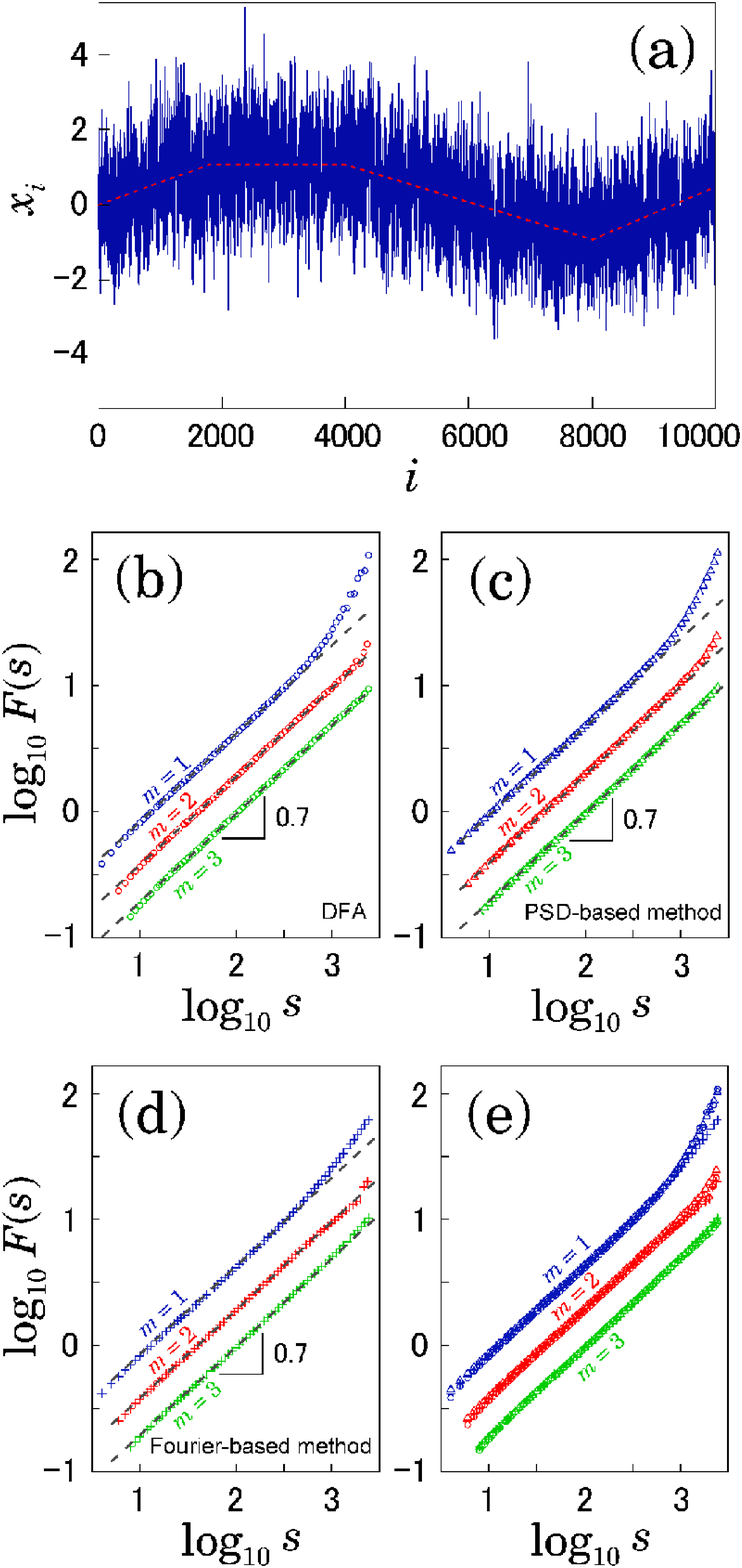}
               \caption{(Color online) Comparison of analysis methods. 
(a) An example of numerically generated time series of the fractional Gaussian noise with $\alpha = 0.7$ after adding piecewise linear trends. The piecewise linear trends are described by dashed lines. The length of the time series is $10^5$. (b) Fluctuation functions $F(s)$ estimated by $m$th-order DFA ($m=1, 2, 3$). (c) $F(s)$ estimated by the PSD-based method [Eq.~\ref{PSD2F}]. (c) $F(s)$ estimated by the Fourier-based method [Eq.~\ref{Four2F}]. (d) Superimposition of (b), (c) and (d). Each point of $F(s)$ represents the mean value of 100 samples.}
               \label{fig:num_test1}
       \end{center}
\end{figure} 

\begin{figure}[t]
       \begin{center}
               \includegraphics[width = 1\linewidth]{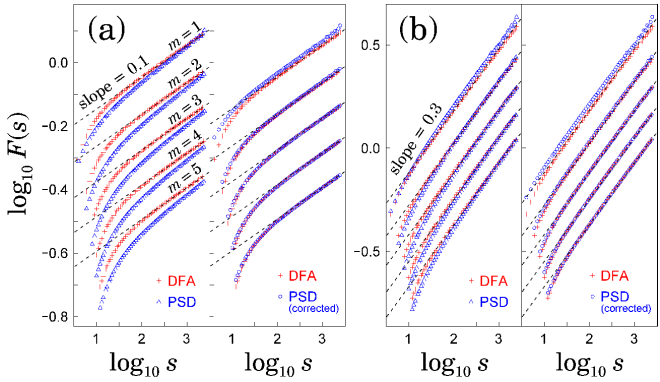}
               \caption{(Color online) Comparison between $m$th-order DFA (+) and PSD-based method with ($\bigcirc $) or without ($\triangle$) the correction factor $c(f_k)$ [Eq.~(\ref{eq:c})]. (a) Analysis of fractional Gaussian noise with $\alpha = 0.1$. (b) Analysis of fractional Gaussian noise with $\alpha = 0.3$. In the left panels of (a) and (b), the correction factor $c(f_k)$ is assumed to be $c(f_k)=1$. Each point of $F(s)$ represents the mean value of 100 samples.}
               \label{fig:num_test2}
       \end{center}
\end{figure} 

\section{Direct connection between DFA and Fourier analysis}\label{sec:relation}

Based on the above mentioned results, the Fourier spectrum of a time series can be converted to the fluctuation function $F(s)$ of DFA. To explain this, we start with the discrete Fourier transform of real-valued time series $\{x_0, x_1, \cdots, x_{N-1} \}$, 
\begin{equation}
X_k := \sum_{i=0}^{N-1} x_i \, e^{-2 \pi j k i /N},  
\end{equation}
where $j$ is the imaginary unit and $k = 0, 1, \cdots, N-1$. In this case, the frequency of a single component indexed by $k$ is given by $f_k = k/N$. If $N$ is odd, the Fourier components with $f_k$ are described by
\begin{equation}
x^{(k)}_i = A_k \cos \left(2 \pi f_k i + \theta_k \right) \quad {\rm for}\ k = 1, 2, \cdots, \frac{N-1}{2}, 
\end{equation}
where $A_k = 2 |X_k|/N$ and $\theta_k = \arg X_k$; if $N$ is even, the Fourier components are described by
\begin{equation}
\left\{ \begin{array}{l l}
x^{(k)}_i = A_k \cos \left(2 \pi f_k i + \theta_k \right) & {\rm for}\ k = 1, 2, \cdots, \frac{N}{2} - 1\\
x^{(k)}_i = A_k \cos \left(\pi i \right) & {\rm for}\ k = N/2
\end{array}
 \right. ,
\end{equation}
where $A_k = 2 |X_k|/N$ and $\theta_k = \arg X_k$ for $k \neq N/2$, and $A_{k} = |X_{k}|/N$ and $\theta_{k} = 0$ for $k=N/2$. Therefore, the time series $\{x_i\}$ can be expressed by the sum of the Fourier components: 
\begin{equation}
x_i = \sum_{k=0}^{\lfloor N/2 \rfloor} x_i^{(k)},
\end{equation}
where $x_i^{(0)} = X_0/n$. 

Using Eqs.~(\ref{eq:Phi2}) and (\ref{eq:c}), the fluctuation function of DFA can be estimated by
\begin{equation}
F (s) = \left[ \frac{1}{\lfloor N/s \rfloor} \sum_{i=1}^{\lfloor N/s \rfloor} \sum_{k=1}^{\lfloor (N-1)/2 \rfloor} \!\!\!\!\!\! c(f_k) \, \Phi^2 (s, i, f_k, A_k, \tilde{\theta}_k) \right]^{1/2}, \label{Four2F}
\end{equation}
where $\lfloor \cdot \rfloor$ is the floor function and $\tilde{\theta}_k = \theta_k + \pi f_k s (2 k -1)$. In the actual calculation of $F(s)$ through Eq.~(\ref{Four2F}), the full information of the phase and amplitude given by the Fourier analysis is required. Therefore, we refer to this as the Fourier-based method.

\begin{figure}[t]
       \begin{center}
               \includegraphics[width = 0.55\linewidth]{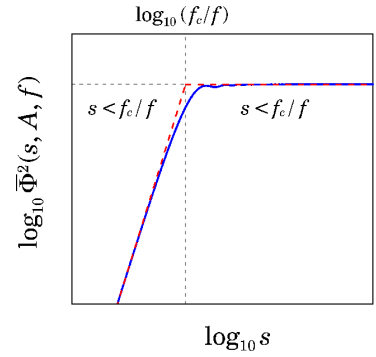}
               \caption{(Color online) Illustration of the single-frequency response function $\overline{\Phi}^2 (s, f, A) $ for a fixed $f$ (solid line) and its approximation [Eq.~(\ref{eq:phi_approx})] (dashed lines).}
               \label{fig:approx_phi}
       \end{center}
\end{figure} 

The single-frequency response function $\overline{\Phi}^2$ [Eq.~(\ref{eq:Phi2ave})] also provides an estimate of $F(s)$ as
\begin{equation}
F (s) = \left[\sum_{k=1}^{\lfloor (N-1)/2 \rfloor} \!\!\!\!\!\! c(f_k) \, \overline{\Phi}^2 (s, f_k, A_k) \right]^{1/2}. \label{PSD2F}
\end{equation}
In this case, calculation of $F(s)$ only require the squared amplitude given by the PSD. Therefore, we refer to this as the PSD-based method.

As shown in Fig.~\ref{fig:num_test1}, if we analyze the time series of the fractional Gaussian noise embedded in piecewise linear trends [Fig.~\ref{fig:num_test1} (a)], both the Fourier-based method and the PSD-based method would be in good agreement with DFA results. However, note that these two methods have different characteristics. In the PSD-based method [Eq.~(\ref{PSD2F})], the detrending procedure is not explicitly included. That is, the $F(s)$ is only given by locally weighted smoothing of the PSD. Therefore, if the trend component forms a large contribution to the total power of the time series, the PSD-based method cannot reproduce the DFA result. On the other hand, in the Fourier-based method [Eq.~(\ref{Four2F})], the detrending procedure is described by the relative location between Fourier phases $\tilde{\theta}_k$ and the local window specified by $i$ (see Fig.~\ref{fig:fig1}). However, a strong trend component in the time series could make the cross terms in Eq.~(\ref{F2decomp}) large. Therefore, the agreement between the DFA and the Fourier-based method is not always excellent. Nevertheless, as we will see, our approach serves to provide a deeper understanding of the methodological features of the DFA. 

It is also important to note that the discretization effect evaluated by $c(f_k)$ [Eq.~(\ref{eq:c})] is remarkably large when a strongly anti-correlated time series with a close-to-zero value of $\alpha$ is analyzed. As shown in the left panel of Fig.~\ref{fig:num_test2} (a), if an anti-correlated signal with $\alpha = 0.1$ is analyzed and the $c(f_k)$ is not considered, the difference between $F(s)$ of the DFA and that of the PSD-based method is clearly seen. The reason for this difference is that, in the PSD of such time series, the power contribution of the high frequency components is much larger than that of the low frequency components. 

\begin{figure}[tb]
       \begin{center}
               \includegraphics[width = 1\linewidth]{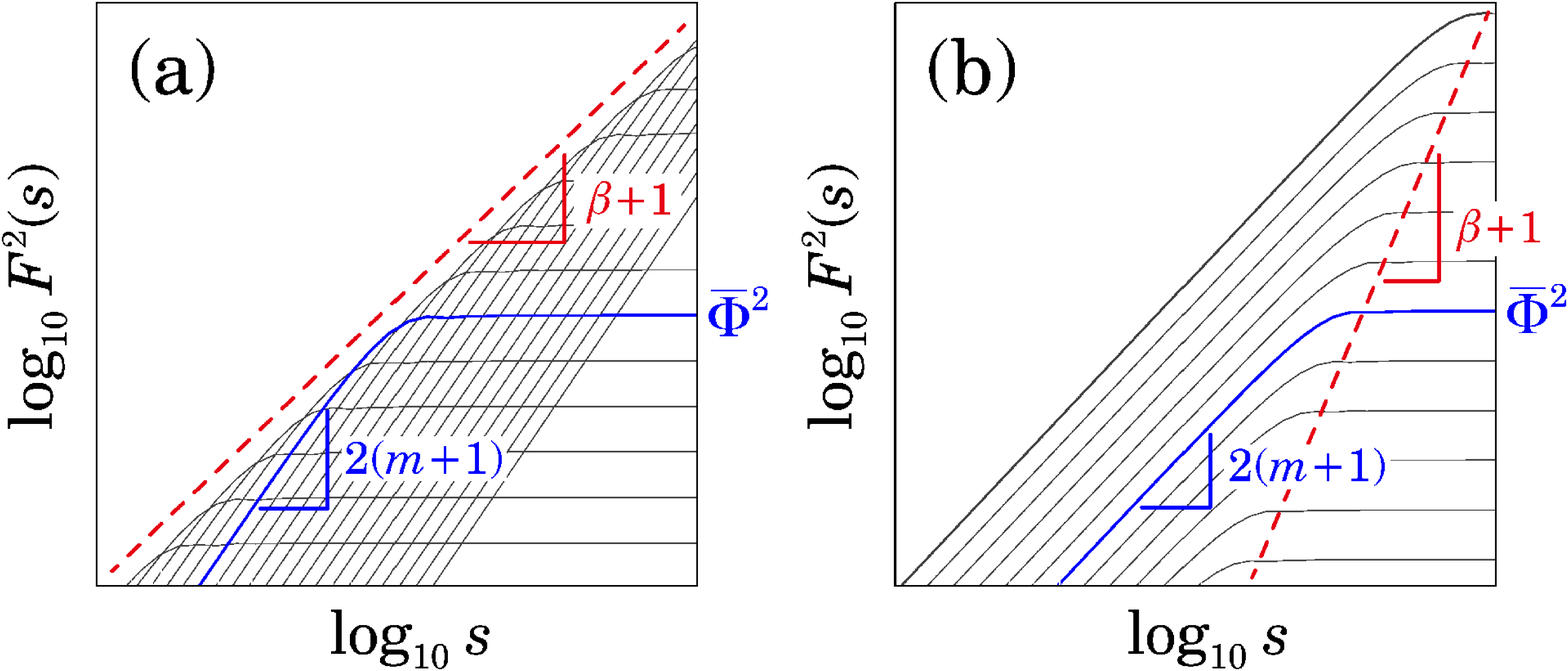}
               \caption{(Color online) Illustration of the squared fluctuation function $F^2(s)$ of DFA. The $F^2(s)$ is given by the weighted sum of single-frequency response functions $\overline{\Phi}^2 (s, f, A(f))$ with different frequencies $f$. Here, the weight is assumed as $A^2(f) \sim f^{-\beta}$. (a) $\beta + 1 < 2 (m + 1)$. (b) $\beta + 1 > 2 (m + 1)$. 
}
               \label{fig:illust_DFAexponent}
       \end{center}
\end{figure}

\begin{figure}[b]
       \begin{center}
               \includegraphics[width = 0.53\linewidth]{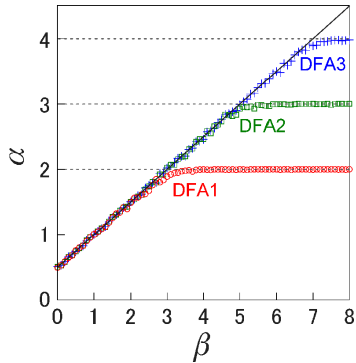}
               \caption{(Color online) Estimated scaling exponent $\alpha$ by DFA versus the scaling exponent $\beta$ of the analyzed time series. Here, numerically generated time series displaying the power-law scaling of the PSD, $S(f) \sim f^{-\beta}$ are analyzed by DFAs. The length of the time series is $N=10^5$. The value of $\alpha$ is estimated by linear regression of $\log_{10} F(s)$ versus $\log_{10} s$ in the range $10^2 \le s \le 10^4$. }
               \label{fig:num_DFA_PSD}
       \end{center}
\end{figure}

\section{Scaling relation between higher-order DFA and PSD} \label{sec:scaling}

Using the frequency-component based description of the DFA [Eq.~(\ref{PSD2F})], we can study the relation between higher-order DFAs and PSD. When $s \ll 1/f$Cthe single-frequency response function $\overline{\Phi}^2$ [Eq.~(\ref{eq:Phi2ave})] can be expanded as
\begin{equation}
\overline{\Phi}^2 (s, f, A) = \frac{ A^2 (\pi f)^{2 m}  s^{2(m+1)}}{8 (2m+3) \prod_{k=0}^{m} (2k+1)^2 } + O \left(s^{2(m+2)}\right), \label{eq:phi_small}
\end{equation}
where $m$ is the order of the least-squares polynomial fit. We have confirmed analytically that Eq.~(\ref{eq:phi_small}) holds when $m = 1, 2, \cdots, 7$. Thus, we conjecture that Eq.~(\ref{eq:phi_small}) holds in general. On the other hand, by taking the limit $s \to \infty$, we obtain
\begin{equation}
\lim_{s \to \infty} \overline{\Phi}^2 (s, f, A) = \frac{A^2}{8 \pi^2 f^2 }. \label{eq:phi_inf}
\end{equation}
Base on the asymptotic behavior of $\overline{\Phi}^2$ [Eqs.~(\ref{eq:phi_small}) and (\ref{eq:phi_inf})], we approximate $\overline{\Phi}^2 (s, f, A)$ by 
\begin{equation}
\overline{\Phi}^2 (s, f, A) \approx  \left\{
\begin{array}{ll}
\displaystyle \frac{ A^2 (\pi f)^{2 m}  s^{2(m+1)}}{8 (2m+3) \prod_{k=0}^{m} (2k+1)^2 } & \ {\rm for \ } s < \displaystyle  \frac{f_c}{f} \\
\\
\displaystyle \frac{A^2 }{8 \pi^2 f^2 } & \ {\rm for \ } s \ge \displaystyle  \frac{f_c}{f} \
\end{array}
\right. , \label{eq:phi_approx}
\end{equation}
where 
\begin{equation}
f_c = \frac{\left\{(2m+3) \prod_{k=0}^{m} (2k+1)^2 \right\}^{1/(2m+2)}}{\pi}.
\end{equation}
This function is illustrated in Fig.~\ref{fig:approx_phi}.

When the PSD function of a stochastic process is given by
\begin{equation}
S_x (f) = \frac{A_0^2}{f^{\beta}}
\end{equation}
in the range $[f_l, 1/2]$, the fluctuation function of the DFA can be evaluated  by assuming Eq.~(\ref{eq:phi_approx}) and $c(f) \approx 1$ as
\begin{widetext}
\begin{eqnarray}
F^2(s) &\sim& \int_{f_l}^{1/2} \! \left. \overline{\Phi}^2 \left(s, f, A \right) \right|_{A^2 = S_x (f)} \, df \label{eq:F2est} \\
&\approx& C^{(1)}_m \int_{0}^{f_c/s} \! f^{2m} s^{2(m+1)} \left( A_0^2\, f^{-\beta}\right) \, df + C^{(2)}_m \int_{f_c/s}^{1/2} \! \frac{1}{f^{2}} A_0^2 f^{-\beta} \, df \nonumber \\
&=& \left( C^{(1)}_m \frac{f_c^{2m-\beta+1}}{2 m - \beta +1}+C^{(2)}_m \frac{f_c^{-\beta-1}}{\beta+1} \right) s^{\beta+1} - C^{(2)}_m \frac{2^{\beta+1}}{\beta + 1} \label{eq:F2calc}
\end{eqnarray}
\end{widetext}
where we set $C^{(1)}_m = \pi^{2 m} / \left\{ 8 (2m+3) \prod_{k=0}^{m} (2k+1)^2 \right\}$ and $C^{(2)}_m = (8 \pi^2)^{-1}$, and assume $f_l \approx 0$ and $\beta < 2m+1$. When $s \gg 1$ and $-1 < \beta < 2m+1$, equation (\ref{eq:F2calc}) results in 
\begin{equation}
F^2(s) \sim s^{\beta+1} \iff F(s) \sim s^{\frac{\beta+1}{2}}. 
\end{equation}
Therefore, for $-1 < \beta < 2m + 1$, we obtain the scaling relation,
\begin{equation}
\alpha = \frac{\beta+1}{2}, \label{eq:scaling_ab}
\end{equation}
where $\alpha$ is the scaling exponent estimated by the DFA as $F(s) \sim s^{\alpha}$. This result shows that the scaling relation [Eq.~(\ref{eq:scaling_ab})] remains independent of the order of the least-squares polynomial fit in the DFA. In other words, the higher-order detrending in the DFA has no adverse effect in the estimation of the scaling exponent.

\begin{figure}[b]
       \begin{center}
               \includegraphics[width = 0.8\linewidth]{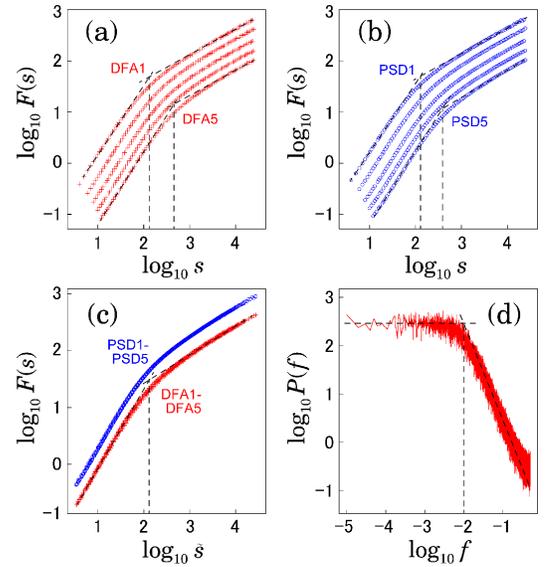}
               \caption{(Color online) Estimation of crossover location in a first-order autoregressive process [Eq.~(\ref{eq:AR1})] with $a = 0.9391014$. Samples of numerically generated time series with length $10^5$ were analyzed. (a) Fluctuation functions $F(s)$ estimated by $m$th-order DFA ($m=1,2, \cdots, 5$). (b) $F(s)$ estimated by the PSD-based method [Eq.~\ref{PSD2F}] with $m=1,2, \cdots, 5$. The plots in (a) and (b) were vertically shifted for improved visibility. (c) $F(s)$ after adjustment by the corrected time scale $\tilde{s} = \log_{10} s - \log_{10} \bar{r}$, where the $\log_{10} \bar{r}$ is calculated under the condition of $S(f) \sim f^{-1}$ (Table \ref{Table1}). The $F(s)$ estimated by the PSD-based method were vertically shifted for improved visibility. (d) The power spectral density (PSD). The analytical prediction of the crossover point is $f_c = -2.0$. Each point represents the mean value of 100 samples. }
               \label{fig:AR_scale}
       \end{center}
\end{figure} 

\section{Limitation of detectable scaling exponent}\label{sec:ub}

In the previous section, the scaling relation between DFA and PSD scaling exponents are derived under the condition of $-1 < \beta < 2m + 1$. This condition implies the existence of the limitation of the detectable scaling exponent by the DFA. When $\beta < -1$ and $s \gg 0$, the last term in Eq.~(\ref{eq:F2calc}) is dominant. This indicates $\alpha = 0$, because of $F^2 (s) \approx {\rm constant} \sim s^{0}$. Therefore, the lower limit of the detectable scaling exponent $\alpha$ is zero. 

On the other hand, if $\beta > 2m+1$, the low frequency components near the lower limit of the scaling range, $f_l$, are dominant in the contribution to the fluctuation function $F(s)$. Therefore, if $2m -\beta + 1$ takes a negative value, and $f_l$ is sufficiently small, $F^2(s)$ given by Eq.~(\ref{eq:F2est}) can be evaluated as
\begin{eqnarray}
F^2(s) &\sim& C^{(1)}_m \int_{f_l}^{f_c/s} \! f^{2m} s^{2(m+1)} \left( A_0^2\, f^{-\beta}\right) \, df \nonumber \\
&=& \left( A_0^2 \, C^{(1)}_m \left[ \frac{f^{2m-\beta + 1}}{2 m - \beta + 1} \right]_{f = f_l}^{f_c/s} \right) s^{2(m+1)} \nonumber \\
&\approx& \left( A_0^2 \, C^{(1)}_m \left[ \frac{f^{2m-\beta + 1}}{2 m - \beta + 1} \right]_{f = f_l}^{f_l + \epsilon } \right) s^{2(m+1)} \nonumber \\
&\sim& s^{2(m+1)}, \label{eq:F2upper}
\end{eqnarray}
where $0 < \epsilon \ll f_c/s$ is chosen such that 
\[
\int_{f_l+\epsilon}^{f_c/s} f^{2m-\beta} \, df \ll \int_{f_l}^{f_l+\epsilon} f^{2m-\beta} \, df. 
\]
Thus, the DFA scaling exponent is given by $\alpha = m+1$. This result shows that the upper limit of the detectable scaling exponent $\alpha$ by the DFA is equal to $m + 1$, which depends only on the order of polynomial detrending, not on $\beta$. In short, the upper limit of the detectable scaling exponent is determined by the power-law exponent of the asymptotic behavior in the left tail of the $\overline{\Phi} (s, f, A)$ for $s \ll 1/f$. This property can be understood intuitively based on the single-frequency response of $m$th-order DFA. As illustrated in Fig.~\ref{fig:illust_DFAexponent}, the left tail of $\overline{\Phi}^2 (s, f, A)$ shows asymptotic power-law behavior as $\overline{\Phi}^2 (s, f, A) \sim s^{2(m+1)}$ [Eq.~(\ref{eq:phi_small})], which corresponds to $F^2(s) \sim s^{2(m+1)}$. As illustrated in Fig.~\ref{fig:illust_DFAexponent}(a), if $\beta+1 < 2 (m + 1)$, the slope $\beta+1$ of $\log_{10} F^2(s)$ versus $\log_{10} s$ can be described by the weighted sum of $\overline{\Phi}^2 (s, f, A(f))$ with different frequencies under the condition of $A^2(f) \sim f^{-\beta}$. On the other hand, as in Fig.~\ref{fig:illust_DFAexponent}(b), if $\beta+1 > 2 (m + 1)$, the slope $\beta+1$ cannot be described by the superposition of $\overline{\Phi}^2 (s, f, A(f))$ under the condition of $A^2(f) \sim f^{-\beta}$. In this case, the slope of $\log_{10} F^2(s)$ versus $\log_{10} s$ is determined by the power-law exponent $2 (m + 1)$ in the left tail of the $\overline{\Phi}^2 (s, f, A)$ for $f \approx f_l$. As shown in Fig.~\ref{fig:num_DFA_PSD}, this property is confirmed by numerical experiments.

\begin{figure}[t]
       \begin{center}
               \includegraphics[width = 0.8\linewidth]{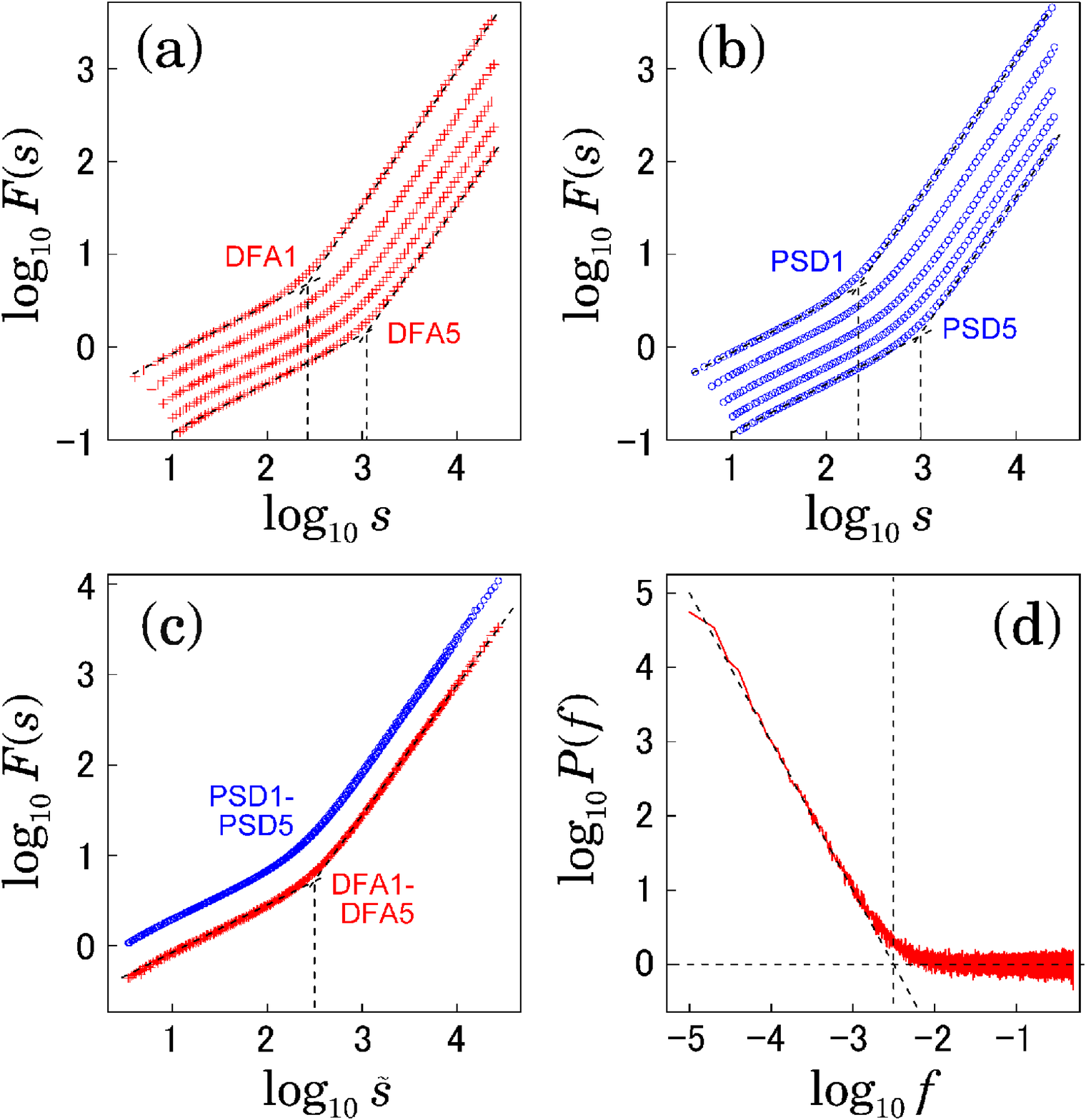}
               \caption{(Color online) Estimation of crossover location in a time series described by the superposition of white noise and Brownian noise [Eq.~(\ref{eq:wB})]. Samples of numerically generated time series with length $10^5$ were analyzed. (a) Fluctuation functions $F(s)$ estimated by $m$th-order DFA ($m=1,2, \cdots, 5$). (b) $F(s)$ estimated by the PSD-based method [Eq.~\ref{PSD2F}] with $m=1,2, \cdots, 5$. The plots in (a) and (b) were vertically shifted for improved visibility. (c) $F(s)$ after adjustment by the corrected time scale $\tilde{s} = \log_{10} s - \log_{10} \bar{r}$, where the $\log_{10} \bar{r}$ is calculated under the condition of $S(f) \sim f^{-1}$ (Table \ref{Table1}). The $F(s)$ estimated by the PSD-based method were vertically shifted for improved visibility. (d) The power spectral density (PSD). The analytical prediction of the crossover point is $f_c = -2.5$. Each point represents the mean value of 100 samples. }
               \label{fig:wB}
       \end{center}
\end{figure}

\begin{table}[htb] 
\caption{Analytical calculations of Eq.~(\ref{eq:a_log10f}). Under the conditions shown in this table, the exact results are expressed by 
$(a - \gamma - \ln 2)/\ln 10 - \log_{10} s$, where $\gamma$ is the Euler-Mascheroni constant. The values of $a$ are summarized in this table. }
\label{Table1}
\begin{center}
\begin{tabular}{c | c c c}
\hline\hline
 &\ $S(f) \sim f^0$ \  &\  $S(f) \sim f^{-1}$ \  &\  $S(f) \sim f^{-2}$\  \\
\hline
DFA1\ & $91/30$ & $9/4$ & $599/420$ \\
DFA2 & $737/210$ & $ 23/8$ & $3197/1260$ \\
DFA3 & $2407/630$ & $129/40$ & $41047/13860$ \\
DFA4 & $28097/6930 $ & $139/40$ & $ 583507/180180$ \\
DFA5 &\ $382201/90090$ & $257/70$ & $ 621259/180180$ \\
\hline
\end{tabular}
\end{center}
\end{table}

\begin{table}[htb] 
\caption{Evaluation of the scale distortion in DFA. The values of $\log_{10} \bar{r}$ are summarized in this table. }
\label{Table2}
\begin{center}
\begin{tabular}{c | c c c}
\hline\hline
 &\ $S(f) \sim f^0$ \  &\  $S(f) \sim f^{-1}$ \  &\  $S(f) \sim f^{-2}$\  \\
\hline
DFA1 & $0.268498$ & $-0.071699$ & $-0.429475$ \\
DFA2 & $0.475305$ & $ 0.199735$ & $ 0.053075$ \\
DFA3 & $0.610419$ & $ 0.351738$ & $ 0.237321$ \\
DFA4 & $0.711943$ & $ 0.460312$ & $ 0.357587$ \\
DFA5 & $0.793605$ & $ 0.545620$ & $ 0.448582$ \\
\hline
\end{tabular}
\end{center}
\end{table}

\section{Scale distortion in DFA}\label{sec:dist}

It is worth to point out that the relation between the time scale in DFA and the corresponding frequency scale in PSD is distorted depending on the order of DFA and on the PSD shape. In the previous studies based on numerical experiments, it has been reported that a deviation of the crossover position observed in the DFA from the corresponding frequency in the PSD is enlarged, as the order of the DFA increases. However, the relation between the scale deviation and the PSD shape has not been pointed out. 

A time scale $s$ in DFA and the corresponding frequency $f$ in PSD would be related as $s = r/f$ or $\log_{10} s = \log_{10} r - \log_{10} f$, in which $r \neq 1$ indicates the scale distortion in the DFA. To evaluate $r$, we calculate
\begin{equation}
\log_{10} \overline{r} = \left\langle \log_{10} f(s) \right\rangle + \log_{10} s, \label{eq:logrm}
\end{equation}
where we define $\left\langle \log_{10} f(s) \right\rangle$ using the single-frequency response function $\overline{\Phi}^2$ as
\begin{equation}
\left\langle \log_{10} f(s) \right\rangle = \frac{\displaystyle \int_{0}^{1/2} \! \left(\log_{10} f \right) \, \overline{\Phi}^2 \left(s, f, A(f) \right)\, df }{\displaystyle \int_{0}^{1/2} \! \overline{\Phi}^2 \left(s, f, A(f) \right)\, df}. \label{eq:log10f}
\end{equation}
In Eq.~(\ref{eq:log10f}), the amplitude spectrum $A(f)$ is determined by the PSD of time series, $S(f)$, as $A(f) \sim \sqrt{S(f)}$. By the relation $f \sim 1/s$, the asymptotic behavior of $\overline{\Phi}^2 \left(s, f, A(f) \right)$ as a function of $f$ is given by Eq.~(\ref{eq:phi_approx}). 

\begin{figure*}[tb]
       \begin{center}
               \includegraphics[width = 0.65\linewidth]{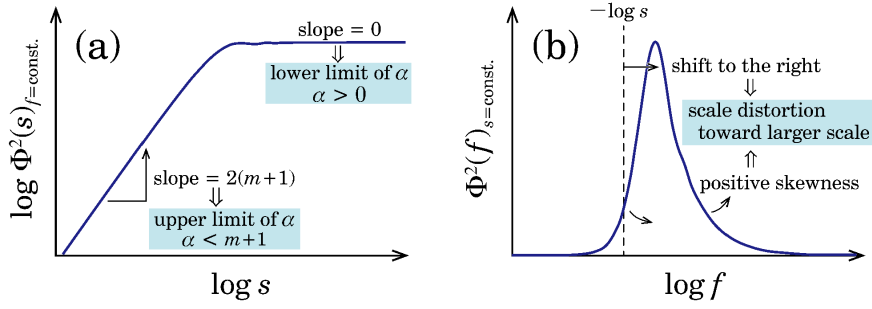}
               \caption{(Color online) Illustration of a single-frequency response functions $\overline{\Phi}^2 (s, f, A)$. (a) $\overline{\Phi}^2$ for a fixed $f$ in $m$th-order DFA. The upper limit of the detectable scaling exponent $\alpha$ is determined by the power-law exponent in the left tail of $\overline{\Phi}^2(s)$.  (b) $\overline{\Phi}^2$ for a fixed $s$. The deviation of the peak position from $- \log s$ and the skewed shape result in the scale distortion in the DFA. }
               \label{fig:illust_phi2}
       \end{center}
\end{figure*} 

Here, assuming $s \gg 1$, which means that the peak position of $\overline{\Phi}^2$ as a function of $f$ is located near $f = 1/s \ll 1$, we approximate Eq,~(\ref{eq:log10f}) as
\begin{equation}
\left\langle \log_{10} f(s) \right\rangle \approx \frac{\displaystyle \int_{0}^{\infty} \! \left(\log_{10} f \right) \, \overline{\Phi}^2 \left(s, f, A(f) \right)\, df }{\displaystyle \int_{0}^{\infty} \! \overline{\Phi}^2 \left(s, f, A(f) \right)\, df}. \label{eq:a_log10f}
\end{equation}
Under some conditions, equation (\ref{eq:a_log10f}) can be calculated analytically. For instance, in the case of second-order DFA and when $S(f) \sim f^0$, we obtain
\begin{equation}
\left\langle \log_{10} f(s) \right\rangle = \frac{737}{210\, \ln 10} - \frac{\gamma + \ln 2 \pi}{\ln 10} - \log_{10} s, 
\end{equation}
where $\gamma \approx 0.57721566$ is the Euler-Mascheroni constant. Furthermore, under some other conditions, the analytical results of Eq.~(\ref{eq:a_log10f}) are expressed by
\begin{equation}
\left\langle \log_{10} f(s) \right\rangle = \frac{a - \gamma - \ln 2 \pi }{\ln 10} - \log_{10} s, 
\end{equation}
where $a$ is a constant. The values of $a$ are summarized in Table \ref{Table1}.

Based on the calculation of $\log_{10} \overline{r}$ [Eq.~(\ref{eq:logrm})], we define the corrected time scale $\tilde{s}$ in DFA as
\begin{equation}
\log_{10} \tilde{s} = - \left\langle \log_{10} f(s) \right\rangle = \log_{10} s - \log_{10} \overline{r}. \label{eq:r_tilde} 
\end{equation}
In the conditions used in Table \ref{Table1}, the values of $\log_{10} \overline{r}$ are given by $(\gamma + \ln 2 \pi - a)/ \ln 10$ (Table \ref{Table2}). 
Note that under the assumption of $S(f) \sim f^{-\beta}$ and $s \gg 1$, the $\log_{10} \overline{r}$ can be approximated by a constant value, but in a more general situation, the $\log_{10} \overline{r}$ is a function of the scale $s$. As seen in Table \ref{Table2}, the $\log_{10} \overline{r}$ increases, as the order of DFA increases, and as the scaling exponent $\beta$ decreases. In general, as the order of DFA increases, the $\log_{10} \tilde{s}$ are shifted to a smaller scale. This shift can be intuitively understood by the shape of $\overline{\Phi}^2 (s, f, A)$ for a fixed $s$. That is, as shown in Fig.~\ref{fig:phi2}(b), in the plot of $\overline{\Phi}^2 (T=1, f, A=1)$ as a function of log frequency, the deviation of its peak position from $f=1/T=1$ and its skewed shape are causes for the scale distortion in the DFA.

When we study the crossover phenomena of the scaling behavior, it is important to consider the scale distortion in DFA. To discuss this, we study two examples displaying crossover phenomena. The first example is a time series $\{x_i\}$ of a first-order autoregressive process (AR1): 
\begin{equation}
x_i = a \,  x_{i-1} + \epsilon_i, \label{eq:AR1}
\end{equation}
where $0 < a < 1$ is the constant parameter and $\epsilon_i$ is white Gaussian noise with zero mean and variance $\sigma^2$. In this process, the PSD is given by
\begin{equation}
P(f) = \frac{\rho f_c}{\pi ( f_c^2 + f^2) }, 
\end{equation}
where 
\begin{equation}
f_c = - \frac{\ln a}{2 \pi},
\end{equation}
and 
\begin{equation}
\rho = \frac{\sigma^2}{1 - a^2}.
\end{equation}
The frequency dependence of this PSD is nearly constant, as it is for white noise, for $f < f_{\rm c}$, and is asymptotically proportional to $f^{-2}$ for $f > f_{\rm c}$. 

Here, by setting $f_c = 10^{-2}$, $\sigma^2 = 1$ and $a=\exp(-2 \pi f_c) = 0.9391014$ in the AR1 process [Eq.~(\ref{eq:AR1})], samples of the time series are numerically generated, and analyzed by DFA. As shown in Fig.~\ref{fig:AR_scale}(a), the locations of crossover points observed in the plots of $\log_{10} F(s)$ versus $\log_{10} s$ are shifted to the right, as the order of DFA increases. Moreover, the location of the crossover point in DFA5 is observed at $\log_{10} s \approx 2.6$, which shows a large deviation from the corresponding crossover point $\log_{10} s = \log_{10} (1/f_c) = 2.0$. To reduce this deviation, we can use the corrected time scales [Eq.~(\ref{eq:r_tilde})]. Here, the value of $\log_{10} \overline{r}$ calculated by assuming $S(f) \sim f^{-1}$, shown in Table \ref{Table2}, is used as an approximation of $\log_{10} \overline{r}$ [Eq.~(\ref{eq:logrm})]. As shown in Fig.~\ref{fig:AR_scale}(c), if the corrected time scales are used, all curves for different order DFAs are almost collapsed into a single curve, and provide a more accurate estimate of the crossover point corresponding to the analytical prediction.

The second example is a time series described by the superposition of white noise and Brownian noise. A Brownian-noise time series $\{x^{\rm (B)}_i \}$ is obtained as the integral of a white-noise time series. When the variance of a white-noise time series is $\sigma_0^2$, the PSD of the integrated series is
\begin{equation}
S_{\rm B} (f) = \frac{\sigma_0^2}{(2 \pi f)^2}. 
\end{equation}
Using this integrated series and another white-noise time series $\{x^{\rm (w)}_i \}$ with variance $\sigma^2$, we generate the time series $\{ x_i \}$ given by
\begin{equation}
x_i = x^{\rm (w)}_i + x^{\rm (B)}_i. \label{eq:wB}
\end{equation}
In this process, the PSD of $\{ x_i \}$ is given by
\begin{equation}
S (f) = \sigma^2 + \frac{\sigma_0^2}{(2 \pi f)^2}, 
\end{equation}
and its crossover frequency $f_c$ is
\begin{equation}
f_c = \frac{\sigma_0}{2 \pi \sigma}.
\end{equation}
The frequency dependence of this PSD is asymptotically proportional to $f^{-2}$ for $f < f_{\rm c}$, and is nearly constant ($\sim f^{0}$) for $f > f_{\rm c}$. 

Here, by setting $f_c = 10^{-2.5}$, $\sigma^2 = 1$, and $\sigma_0 = 2 \pi \sigma f_c = 0.01986918$, samples of the time series are numerically generated and analyzed by DFA. As shown in Fig.~\ref{fig:wB},  the crossover points are observed to shift as in the previous example. In contrast, as shown in Fig.~\ref{fig:wB}, if we use the corrected time scale $\log_{10} \tilde{s}$, all curves for different order DFAs are almost collapsed into a single curve, and provide a more accurate estimate of the crossover point corresponding to the analytical prediction $- \log_{10} f_c = 2.5$. 

The above results demonstrate that the corrected time scales $\log_{10} \tilde{s}$ based on $\log_{10} \overline{r}$ calculated by assuming $S(f) \sim f^{-1}$ can improve the crossover point estimation when the observed scaling exponents lie in the range $0 \le \beta \le 2$ (or $0.5 \le \alpha \le 1.5$). In more general cases, the corrected time scales can be estimated by Eq.~(\ref{eq:log10f}). 

\begin{figure}[tb]
       \begin{center}
               \includegraphics[width = 1\linewidth]{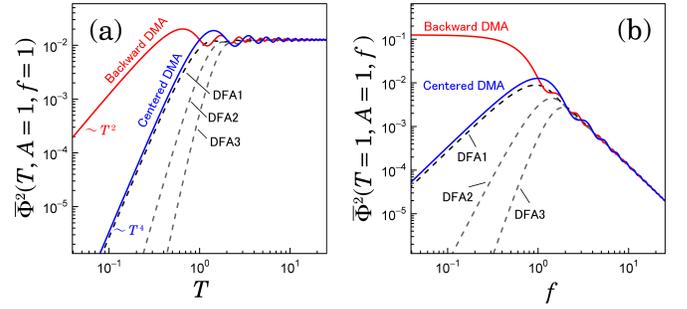}
               \caption{(Color online) Single-frequency response functions $\overline{\Phi}^2 (T, f, A)$ of backward detrending moving average algorithm (BDMA) and centered DMA (CDMA) methods, compared with $m$th order DFA ($m=1, 2, 3$). (a) $\overline{\Phi}^2 (T, f=1, A=1)$ versus $T$. (c) $\overline{\Phi}^2 (T=1, f, A=1)$ versus $f$. }
               \label{fig:MA_methods}
       \end{center}
\end{figure} 

\section{Summary and discussion}\label{sec:summary}

In this work, we have derived the direct connection between higher-order DFA and Fourier analysis using the single-frequency response of the DFA [Eqs.~(\ref{Four2F}) and (\ref{PSD2F})], and shown the usefulness of our approach for understanding the methodological features of DFA. As an important result, it is confirmed that, when we analyze stochastic processes displaying a power-law scaling of the PSD, $S(f) \sim f^{-\beta}$, the higher-order detrending in the DFA has no adverse effect on the estimation of the DFA scaling exponent $\alpha$ which satisfies the scaling relation $\alpha = (\beta+1)/2$. However, there exists the limitation of the detectable scaling exponent $\alpha$, $0 < \alpha < m + 1$, where $m$ is the order of the polynomial fit used in the DFA. This fact suggests that, in the case of the first-order DFA, the observation of scaling behavior with $\alpha \approx 2$ may be a spurious detection. Therefore, in such a case, careful testing is required. It is also important to note the scale distortion which is present in DFA and which we suggest can be reduced by employing the corrected time scale in DFA. As illustrated in Fig.~\ref{fig:illust_phi2}, the above-mentioned features are clearly understood by the shape of the single-frequency response function.

Our approach based on the single-frequency response can provide a more general framework for understanding the methodological features of the variants of DFA using different types of detrending. For instance, we can consider the DFA using a simple moving average filtering, known as the detrending moving average algorithm (DMA) \cite{carbone2004analysis,alvarez2005detrending,xu2005quantifying}. In the case of the backward DMA (BDMA) method, the moving average for a window with the size $s$ is defined as
\begin{equation}
\bar{y}^{(s)}_k = \frac{1}{n} \sum_{j=0}^{s-1} y_{k-j},
\end{equation}
where $\{y_k\}$ is the integrated time series defined by Eq.~(\ref{eq:integ_x}). Alternatively, in the case of the centered DMA (CDMA) method, the moving average for a window with the size $s$ is defined as
\begin{equation}
\bar{y}^{(s)}_k = \frac{1}{n} \sum_{j=-\lfloor s/2 \rfloor}^{\lfloor (s-1)/2 \rfloor} y_{k+j}. 
\end{equation} 
where $\{y_k\}$ is the integrated time series defined by Eq.~(\ref{eq:integ_x}). Using these moving averages $\{\bar{y}^{(s)}_k\}$, the fluctuation function $F(s)$ is calculated by the root mean square deviation of the integrated time series around the moving average $\{\bar{y}^{(s)}_k\}$. 

To study the methodological properties of the BDMA and CDMA methods, we can calculate the single-frequency response function $\overline{\Phi}^2 (T, f, A)$ of these methods. These exact expressions of $\overline{\Phi}^2 (T, f,  A)$ are given by
\begin{widetext}
\begin{equation}
\overline{\Phi}^2 (T, f, A) = \frac{A^2 \left\{ 2 \pi ^2 f^2 T^2 - 2 \pi  f T \sin (2 \pi  f T) - \cos (2 \pi  f T) + 1\right\}}{16 \pi ^4 f^4 T^2}, \label{eq:bma}
\end{equation}
\end{widetext}
and by
\begin{equation}
\overline{\Phi}^2 (T, f, A) = \frac{A^2 \left\{ \sin (\pi  f T)-\pi  f T \right\}^2}{8 \pi ^4 f^4 T^2}, \label{eq:cma}
\end{equation}
respectively. Figure \ref{fig:MA_methods} shows the $\overline{\Phi}^2 (T, f, A)$ of the BDMA and CDMA methods, compared with standard DFA. The plot of the $\overline{\Phi}^2 (T, A, f)$ shows similarity between CMA and first-order DFA, which indicates that both methods have a similar frequency response property. As discussed in section \ref{sec:ub}, the power-law tail structure of $\overline{\Phi}^2 (T, f, A)$ for $s \ll 1/f$ determines the upper limit of the detectable scaling exponent by DFA. When $s \ll 1/f$Cequations (\ref{eq:bma}) and (\ref{eq:cma}) are expanded as
\begin{equation}
\overline{\Phi}^2 (T, f, A) = \frac{A^2 T^2}{8} + O (T^4), 
\end{equation}
and as
\begin{equation}
\overline{\Phi}^2 (T, f, A) = \frac{A^2 \pi^2 f^2 T^4}{288} + O (T^6), 
\end{equation}
respectively. Therefore, scaling exponents $\alpha$ estimated by these methods are bounded by $1$ and $2$, respectively.

Our analytical approach is applicable to higher-order DFAs and their variants, and can contribute to a deeper understanding of the methodological properties of the DFA. Our approach could provide valuable information for the improvement of the DFA methodology. 

\section*{Acknowledgements}

The author would like to thank Professors Taishin Nomura and Yasuyuki Suzuki for fruitful comments. This work was supported by JSPS KAKENHI Grant Number 15K01285.

\appendix
\section{Decomposition of least-squares polynomial fits}

Consider square-integral functions $x(t)$, $x^{(1)} (t)$ and $x^{(2)} (t)$, and assume 
\begin{equation}
x(t) = x^{(1)} (t)+x^{(2)} (t) \label{xx12}
\end{equation}
in the range $\left[t_1, t_2 \right]$. We denote $k$th-order least-squares polynomials to $x(t)$, $x^{(1)} (t)$ and $x^{(2)} (t)$ by $p(t)$C$p^{(1)} (t)$ and $p^{(2)} (t)$, respectivly, which are described by 
\begin{eqnarray}
p(t) &=& \sum_{k = 0}^{m} a_k \, t^k, \label{p_ak}\\
p^{(i)}(t) &=& \sum_{k = 0}^{m} a^{(i)}_k t^k \quad (i = 1, 2),
\end{eqnarray}
where $\{a_k\}$ and $\{a^{(i)}_k\}$ ($i = 1, 2$) are constant. In this case, the following holds: 
\begin{equation}
p(t) = p^{(1)}(t) + p^{(2)}(t). \label{fit_decomp}
\end{equation}

The proof of Eq.~(\ref{fit_decomp}) is the following. The coefficients $\{a_k\}$ in Eq.~(\ref{p_ak}) are given by minimization of the sum of squared vertical residuals, 
\begin{equation}
I(\{a_k\}) = \int_{t_1}^{t_2} \!\! \left\{f(t) - p(t) \right\}^2\, dt. 
\end{equation}
Therefore, $\{a_k\}$ satisfies 
\begin{equation}
\frac{\partial I(\{a_k\})}{\partial a_k} = 0 \quad (k=0, 1, \cdots, m), 
\end{equation}
which results in
\begin{equation}
\int_{t_1}^{t_2} \!\! \left\{f(t) - p(t) \right\} t^k \, dt = 0. \label{eq_fit_k}
\end{equation}
In addition, $\{a^{(1)}_k\}$ and $\{a^{(2)}_k\}$ satisfy
\begin{eqnarray}
\int_{t_1}^{t_2} \!\! \left\{f^{(1)}(t) - p^{(1)}(t) \right\} t^k \, dt &=& 0, \label{p1} \\
\int_{t_1}^{t_2} \!\! \left\{f^{(2)}(t) - p^{(2)}(t) \right\} t^k \, dt &=& 0. \label{p2}
\end{eqnarray}
From Eqs.~(\ref{eq_fit_k}), (\ref{p1}) and (\ref{p2}), we can obtaion
\begin{eqnarray}
\int_{t_1}^{t_2} \!\! \left\{p (t) - \left({p}^{(1)}(t) + {p}^{(2)}(t)\right) \right\} t^k \, dt &=& 0 \nonumber \\
\sum_{i = 0}^{m} \left\{a_i-\left(a^{(1)}_i+a^{(2)}_i \right)\right\} 
\int_{t_1}^{t_2} \!\! t^{i+k} \, dt &=& 0 
\end{eqnarray}
This equation must always hold true for $k = 1, 2, \cdots, m$. Therefore, we obtain $a_i = a^{(1)}_i+a^{(2)}_i $ ($i=1, \cdots, m$), i.e. Eq.~(\ref{fit_decomp}).

\begin{widetext}

\section{Analytical formulas of $\Phi^2$ and $\overline{\Phi}^2$}
\subsection{Zeroth-order detrending ($m=0$)}
\begin{eqnarray}
\Phi^2 (T, f, A, \theta) &=& \frac{A^2 \left\{ \cos (2 \theta ) \left(2 \sin ^2(\pi  f T)-\pi  f T \sin (2 \pi  f T)\right)+\cos (2 \pi  f T)+ 2 \pi ^2 f^2 T^2 -1\right\}}{16 \pi ^4 f^4 T^2} \\
\overline{\Phi}^2 (T, f, A) &=& \frac{A^2 \left(2 \pi ^2 f^2 T^2+\cos (2 \pi  f T)-1\right)}{16 \pi ^4 f^4 T^2}
\end{eqnarray}

\subsection{First-order detrending ($m=1$)}
\begin{eqnarray}
\Phi^2 (T, f, A, \theta) &=& \frac{A^2}{32 \pi ^6 f^6 T^4} \Big\{\left(3-4 \pi ^2 f^2 T^2\right) \cos (2 \pi  f T-2 \theta )-6 \cos (2 \theta )-4 \pi ^2 f^2 T^2 \cos (2 \theta ) \nonumber \\
&& + 3 \cos (2 (\pi  f T+\theta ))-4 \pi ^2 f^2 T^2 \cos (2 (\pi  f T+\theta ))+12 \pi  f T \sin (2 \pi  f T) \nonumber \\
&& + 6 \pi  f T \sin (2 \pi  f T-2 \theta )-\pi ^3 f^3 T^3 \sin (2 \pi  f T-2 \theta ) + 6 \pi  f T \sin (2 (\pi  f T+\theta )) \nonumber \\
&& - \pi ^3 f^3 T^3 \sin (2 (\pi  f T+\theta ))  -6-8 \pi ^2 f^2 T^2+4 \pi ^4 f^4 T^4+\left(6-4 \pi ^2 f^2 T^2\right) \cos (2 \pi  f T) \Big\} \\
\overline{\Phi}^2 (T, f, A) &=& \frac{A^2}{16 f^6 \pi ^6 T^4} \left\{2 \pi^4 f^4 T^4-4 \pi^2 f^2 T^2 -3 + \left(3-2 \pi^2 f^2 T^2\right) \cos (2 \pi f T) + 6 \pi f T \sin (2 \pi f T) \right\}. 
\end{eqnarray}

\subsection{Second-order detrending ($m=2$)}
\begin{eqnarray}
\Phi^2 (T, f, A, \theta) &=& \frac{A^2}{32 \pi ^8 f^8 T^6}
 \big\{ 6 \left(\pi ^4 f^4 T^4-24 \pi ^2 f^2 T^2+15\right) \cos (2 \pi  f T)+90 \cos (2 \theta ) + 4 \pi ^6 f^6 T^6-18 \pi ^4 f^4 T^4 \nonumber \\
&& + \left(-9 \pi ^4 f^4 T^4+78 \pi ^2 f^2 T^2-45\right) \cos (2 \pi  f T-2 \theta ) + 24 \pi ^2 f^2 T^2 \cos (2 \theta ) -36 \pi ^2 f^2 T^2-90 \nonumber \\
&& +6 \pi ^4 f^4 T^4 \cos (2 \theta )-45 \cos (2 (\pi  f T+\theta )) + 78 \pi ^2 f^2 T^2 \cos (2 (\pi  f T+\theta )) \nonumber \\
&& -9 \pi ^4 f^4 T^4 \cos (2 (\pi  f T+\theta ))+180 \pi  f T \sin (2 \pi  f T) - 48 \pi ^3 f^3 T^3 \sin (2 \pi  f T) \nonumber \\
&& -90 \pi  f T \sin (2 \pi  f T-2 \theta )+36 \pi ^3 f^3 T^3 \sin (2 \pi  f T-2 \theta ) - \pi ^5 f^5 T^5 \sin (2 \pi  f T-2 \theta ) \nonumber \\
&& -90 \pi  f T \sin (2 (\pi  f T+\theta )) + 36 \pi ^3 f^3 T^3 \sin (2 (\pi  f T+\theta ))-\pi ^5 f^5 T^5 \sin (2 (\pi  f T+\theta ))\big\} \\
\overline{\Phi}^2 (T, f, A) &=& \frac{A^2}{16 \pi ^8 f^8 T^6} \big\{ 3 \left(\pi ^4 f^4 T^4-24 \pi ^2 f^2 T^2+15\right) \cos (2 \pi  f T) - 6 \pi  f T \left(4 \pi ^2 f^2 T^2-15\right) \sin (2 \pi  f T) \nonumber \\
&&  + 2 \pi ^6 f^6 T^6 - 9 \pi ^4 f^4 T^4-18 \pi ^2 f^2 T^2-45 \big\}
\end{eqnarray}

\subsection{Third-order detrending ($m=3$)}
\begin{eqnarray}
\Phi^2 (T, f, A, \theta) &=& \frac{A^2}{32 \pi ^{10} f^{10} T^8} \big\{ \left(-8 \pi ^6 f^6 T^6+780 \pi ^4 f^4 T^4-5580 \pi ^2 f^2 T^2+3150\right) \cos (2 \pi  f T) \nonumber \\
&& - \left(16 \pi ^6 f^6 T^6 - 540 \pi ^4 f^4 T^4+ 2880 \pi ^2 f^2 T^2-1575\right) \cos (2 \pi  f T-2 \theta ) - 32 \pi ^6 f^6 T^6 \nonumber \\
&& - 3150 \cos (2 \theta )-540 \pi ^2 f^2 T^2 \cos (2 \theta )-60 \pi ^4 f^4 T^4 \cos (2 \theta ) -120 \pi ^4 f^4 T^4 +4 \pi ^8 f^8 T^8 \nonumber \\
&& - 8 \pi ^6 f^6 T^6 \cos (2 \theta )+1575 \cos (2 (\pi  f T+\theta ))-2880 \pi ^2 f^2 T^2 \cos (2 (\pi  f T+\theta )) \nonumber \\
&& + 540 \pi ^4 f^4 T^4 \cos (2 (\pi  f T+\theta ))-16 \pi ^6 f^6 T^6 \cos (2 (\pi  f T+\theta )) -720 \pi ^2 f^2 T^2 \nonumber \\
&& + 6300 \pi  f T \sin (2 \pi  f T)-2760 \pi ^3 f^3 T^3 \sin (2 \pi  f T)+120 \pi ^5 f^5 T^5 \sin (2 \pi  f T) \nonumber \\
&& + 3150 \pi  f T \sin (2 \pi  f T-2 \theta )-1560 \pi ^3 f^3 T^3 \sin (2 \pi  f T-2 \theta ) -\pi ^7 f^7 T^7 \sin (2 (\pi  f T+\theta )) \nonumber \\
&& + 120 \pi ^5 f^5 T^5 \sin (2 \pi  f T-2 \theta )-\pi ^7 f^7 T^7 \sin (2 \pi  f T-2 \theta ) + 120 \pi ^5 f^5 T^5 \sin (2 (\pi  f T+\theta )) \nonumber \\
&& + 3150 \pi  f T \sin (2 (\pi  f T+\theta ))-1560 \pi ^3 f^3 T^3 \sin (2 (\pi  f T+\theta )) -3150 \big\} \\
\overline{\Phi}^2 (T, f, A) &=& \frac{A^2}{16 \pi ^{10} f^{10} T^8} \{ 30 \pi  f T \left(2 \pi ^4 f^4 T^4-46 \pi ^2 f^2 T^2+105\right) \sin (2 \pi  f T) + 2 \pi ^8 f^8 T^8 - 16 \pi ^6 f^6 T^6 - 60 \pi ^4 f^4 T^4  \nonumber \\
&& - \left(4 \pi ^6 f^6 T^6-390 \pi ^4 f^4 T^4+2790 \pi ^2 f^2 T^2-1575\right) \cos (2 \pi  f T) - 360 \pi ^2 f^2 T^2- 1575 \big\}
\end{eqnarray}

\subsection{Fourth-order detrending ($m=4$)}
\begin{eqnarray}
\overline{\Phi}^2 (T, f, A) &=& \frac{A^2}{16 \pi ^{12} f^{12} T^{10}} \{
2 \pi ^{10} f^{10} T^{10} -25 \pi ^8 f^8 T^8 -150 \pi ^6 f^6 T^6 - 1575 \pi ^4 f^4 T^4 -15750 \pi ^2 f^2 T^2 \nonumber \\
&& + 5 \left(\pi ^8 f^8 T^8-264 \pi ^6 f^6 T^6+7245 \pi ^4 f^4 T^4-36540 \pi ^2 f^2 T^2+19845\right) \cos (2 \pi  f T) \nonumber \\
&& - 30 \pi  f T \left(4 \pi ^6 f^6 T^6-287 \pi ^4 f^4 T^4+3360 \pi ^2 f^2 T^2-6615\right) \sin (2 \pi  f T)-99225\}
\end{eqnarray}

\subsection{Fifth-order detrending ($m=5$)}
\begin{eqnarray}
\overline{\Phi}^2 (T, f, A) &=& \frac{A^2}{16 \pi ^{14} f^{14} T^{12}} \{2 \pi ^{12} f^{12} T^{12}-36 \pi ^{10} f^{10} T^{10}-315 \pi ^8 f^8 T^8-5040 \pi ^6 f^6 T^6-85050 \pi ^4 f^4 T^4 \nonumber \\
&& - 1190700 \pi ^2 f^2 T^2 - 3 \left(2 \pi ^{10} f^{10} T^{10}-1155 \pi ^8 f^8 T^8+81480 \pi ^6 f^6 T^6-1417500 \pi ^4 f^4 T^4 \right. \nonumber \\
&& + \left. 6151950 \pi ^2 f^2 T^2-3274425\right) \cos (2 \pi  f T)+ 210 \pi  f T \left(\pi ^8 f^8 T^8-168 \pi ^6 f^6 T^6 \right. \nonumber \\
&& \left.+5724 \pi ^4 f^4 T^4-51030 \pi ^2 f^2 T^2+93555\right) \sin (2 \pi  f T) - 9823275 \}
\end{eqnarray}

\subsection{Sixth-order detrending ($m=6$)}
\begin{eqnarray}
\overline{\Phi}^2 (T, f, A) &=& \frac{A^2}{16 \pi ^{16} f^{16} T^{14}} \Big\{
2 \pi ^{14} f^{14} T^{14}-49 \pi ^{12} f^{12} T^{12}-588 \pi ^{10} f^{10} T^{10}-13230 \pi ^8 f^8 T^8 \nonumber \\
&& -330750 \pi ^6 f^6 T^6-7640325 \pi ^4 f^4 T^4-137525850 \pi ^2 f^2 T^2-42 \pi  f T \left(8 \pi ^{10} f^{10} T^{10} \right. \nonumber \\
&& \left. -2670 \pi ^8 f^8 T^8+203040 \pi ^6 f^6 T^6-4916835 \pi ^4 f^4 T^4+38045700 \pi ^2 f^2 T^2-66891825\right) \sin (2 \pi  f T) \nonumber \\ 
&& +7 \left(\pi ^{12} f^{12} T^{12}-1104 \pi ^{10} f^{10} T^{10}+162810 \pi ^8 f^8 T^8-6875820 \pi ^6 f^6 T^6+95582025 \pi ^4 f^4 T^4 \right. \nonumber \\
&& \left. -381704400 \pi ^2 f^2 T^2+200675475\right) \cos (2 \pi  f T)-1404728325 \Big\} 
\end{eqnarray}

\subsection{Seventh-order detrending ($m=7$)}
\begin{eqnarray}
\overline{\Phi}^2 (T, f, A) &=& \frac{A^2}{16 \pi ^{18} f^{18} T^{16}} \Big\{
-2 \pi ^{16} f^{16} T^{16}+64 \pi ^{14} f^{14} T^{14}+1008 \pi ^{12} f^{12} T^{12}+30240 \pi ^{10} f^{10} T^{10} \nonumber \\
&& +1039500 \pi ^8 f^8 T^8+34927200 \pi ^6 f^6 T^6+1021620600 \pi ^4 f^4 T^4+22475653200 \pi ^2 f^2 T^2 \nonumber \\ 
&& -126 \pi  f T \left(4 \pi ^{12} f^{12} T^{12}-2380 \pi ^{10} f^{10} T^{10}+346170 \pi ^8 f^8 T^8-17901180 \pi ^6 f^6 T^6 \right. \nonumber \\
&& \left. +358107750 \pi ^4 f^4 T^4-2541889350 \pi ^2 f^2 T^2 +4347968625\right) \sin (2 \pi  f T) \nonumber \\ 
&& +\left(8 \pi ^{14} f^{14} T^{14}-15372 \pi ^{12} f^{12} T^{12}+4169340 \pi ^{10} f^{10} T^{10}-353617110 \pi ^8 f^8 T^8 \right. \nonumber \\ 
&& +11373169500 \pi ^6 f^6 T^6-138684996450 \pi ^4 f^4 T^4+525368393550 \pi ^2 f^2 T^2 \nonumber \\
&& \left.  -273922023375\right) \cos (2 \pi  f T)+273922023375 \Big\} 
\end{eqnarray}

\end{widetext}



\end{document}